\documentclass[twoside]{article}
\usepackage{qic,epsfig}

\newcommand{\beq}{\begin{equation}}
\newcommand{\eeq}{\end{equation}}
\newcommand{\ket}[1]{\left| {#1} \right>}

\usepackage{graphics}

\begin{document}
\setlength{\textheight}{8.0truein}    

\runninghead{Title  $\ldots$}
            {Author(s) $\ldots$}

\normalsize\textlineskip
\thispagestyle{empty}
\setcounter{page}{1}

\copyrightheading{0}{0}{2003}{000--000}

\vspace*{0.88truein}

\alphfootnote

\fpage{1}

\centerline{\bf HOW TO BUILD A 300 BIT, 1 GIGA-OPERATION QUANTUM COMPUTER}
\vspace*{0.37truein}
\centerline{\footnotesize ANDREW M. STEANE}
\vspace*{0.015truein}
\centerline{\footnotesize\it
Centre for Quantum Computation, Department of Atomic and Laser
Physics,}
\centerline{\footnotesize\it
Clarendon Laboratory, Parks Road, Oxford, OX1 3PU,
England}
\baselineskip=10pt
\vspace*{0.225truein}
\publisher{(received date)}{(revised date)}

\vspace*{0.21truein}

\abstracts{
Experimental methods for laser control of trapped ions have reached
sufficient maturity that it is possible to set out in detail a
design for a large quantum computer based on such methods, without
any major omissions or uncertainties. The main features of such a
design are given, with a view to identifying areas for study. The
machine is based on 13000 ions moved via 20$\mu$m vacuum channels
around a chip containing 160000 electrodes and associated classical
control circuits; 1000 laser beam pairs are used to manipulate the
hyperfine states of the ions and drive fluorescence for readout. The
computer could run a quantum algorithm requiring $10^9$ logical
operations on 300 logical qubits, with a physical gate rate of 1 MHz
and a logical gate rate of 8 kHz, using methods for quantum gates
that have already been experimentally implemented. Routes for faster
operation are discussed.
}{}{}

\vspace*{10pt}

\vspace*{3pt}
\communicate{}

\vspace*{1pt}\textlineskip    

The problem of making quantum computing feasible has been
addressed from two main directions: the implementation of
experimental methods to manipulate quantum systems, and the
discovery of general methods to control errors and noise without
destroying the coherence of a quantum computation. Several recent
developments of experimental methods for laser-control of trapped
ions have allowed a much greater degree of confidence that such
methods can be extended to the sort of system size which would
make a significant quantum computer (QC)
\cite{03Leibfried,04Riebe,04Barrett,04Cirac1,05Brickman,05Haffner,05Haljan,05Leibfried,04Chiaverini}.
Meanwhile, quantum error correction (QEC) circuits have been
analyzed in sufficient detail to allow a discussion of the
physical requirements of a large computer without any major
omissions \cite{03Steane,02Steane2}. By bringing these two aspects
together, it is possible to outline one way to build a QC that is
technologically feasible and does not require a very great degree
of uncertainty in extrapolation from already established results.
This article offers such an outline. Although we don't yet know if
various details of the method described here are the best ones to
use to build a large machine, it sets out the main issues and
clarifies the vision for future work.

It is essential to the present argument that I only invoke
micro-fabrication technology that is already available, time- and
distance-scales close to those already explored in laboratory
experiments, and techniques that have been demonstrated
sufficiently to show all the ingredients required for repetitive
QEC working together in a single experiment. Many different QC
designs could be produced if one were to relax one or more of
these requirements, but this would be a less valuable exercise
because unexpected noise sources are always discovered when
experiments are done. The only assumptions I make are that
established techniques can be implemented with more laser power
and stability, in a parallelized form, and that the ion trap
structures can be integrated with the necessary control circuitry.

The basic elements which would be required to make a
general-purpose QC (qubits, gates and readout) were proposed by
Deutsch \cite{85Deutsch}. Discussion of possible physical
implementations has generated a large literature; most of the
major considerations have been summarised by DiVincenzo
\cite{00DiVincenzo}. The essential requirement is that the
computer can implement fault-tolerant QEC on a useful number of
qubits, because once it can do that then it can compute as well.
In the network model of computation which we adopt here, for QEC
the operation needed most frequently is to move quantum
information from one place to another, the next most frequent is a
logic gate such as controlled-not or controlled-phase between
neighbouring qubits, and after that measurement of qubits and
single-qubit rotations\cite{02Steane2}.

\begin{figure}[htbp]
\centerline{\epsfig{file=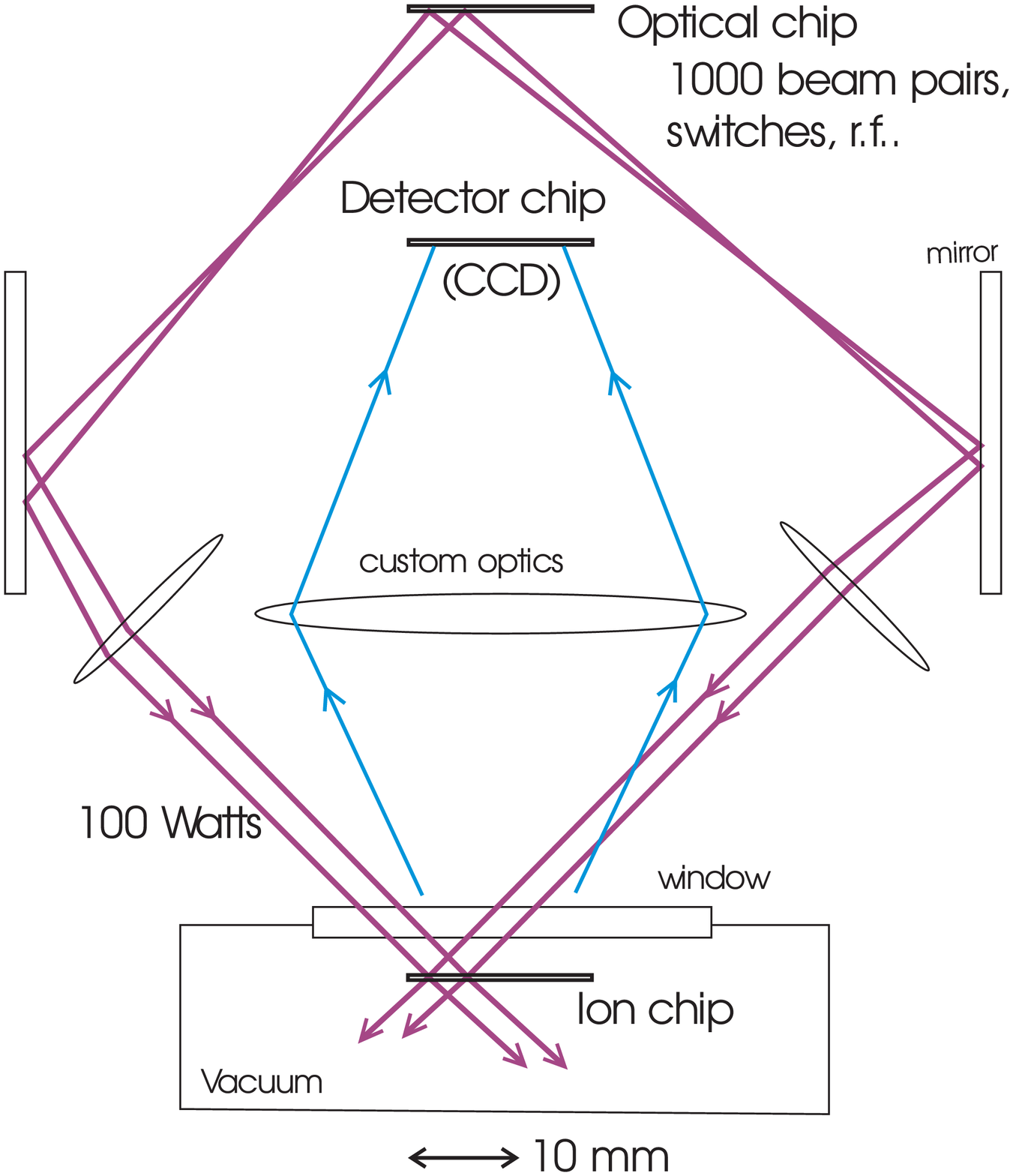, width=8.2cm}} 
\fcaption{Schematic diagram of the complete computer. An optical
chip contains laser sources, optical switches and r.f. control
circuitry for the laser pulses; the multiple laser beams (2 of 1000 pairs
are shown) are imaged onto an `ion chip' (IC) in vacuum, containing the array of ion
traps and the control circuitry for moving ions around. The detector
registers optical fluorescence; its elements could alternatively be incorporated onto
the IC. The optical chip could alternatively be placed inside the vacuum chamber,
close to the IC, or else replaced by conventional methods (stand-alone lasers, cavities and AOM's).}
\end{figure}

\section{Overview}
\noindent
In this section the computer is described.
The reasoning and technical arguments which
lead to the choices of parameter values are given in section 2,
and possibilities for improvements are discussed in section 3.

A schematic diagram of the machine is shown in
figure 1. The main parameters are defined below and listed in
tables 1 and 2.  Table 1
gives parameters of the overall design, such as the QEC encoding,
the noise rates, the gate rate and the overall stability. Table 2
gives more detailed information on the physical processes and
design.

\vspace*{4pt}   
\begin{table}[hb]
\tcaption{The main logical parameters of the computer. The
parameters above the line describe the large-scale encoding. Those
below the line use information from table 2 to provide the timing
and noise parameters. The recovery `crash' probability is the
probability that recovery of a block fails, for example owing to
an uncorrectable error, on the usual uncorrelated stochastic error
model. Since this scales as $\gamma_2^7$ it is a very sensitive
function of the noise level; the important point is that the error
rates listed are low enough to allow a very large number of
logical gates. }
\centerline{\footnotesize\smalllineskip
\begin{tabular}{lll}
Encoding \\
n. ion pairs (p-bits) & $N$         & 6444 \\
block code parameters & \multicolumn{2}{l}{$[[n,k,d]] \;\;\;\;\;\;\;\;\;\;\;\;\; [[127,29,15]]$} \\
ancilla circuit parameters & \multicolumn{2}{l}{$N_A,\; w$ \rule{13ex}{0pt} 1939, 47} \\
data$+$anc. bits per block & $4n+k$ & 537 \\
n. blocks                         & $b = N/(4n+k)$ & 12 \\
n. parallel operations & $N_{\rm P} = 2 b N_A/w$ & 990 \\
n. parallel measurements & $b n$  & 1524 \\
\hline
Overall performance\\
{physical gate time} $\;\;\;\;\tau_g=$ & $\frac{2}{\nu_{\rm spl}} + \frac{10}{\nu_{r}} + \tau_{\rm cool} + \tau_p$ & $1.2 \,\mu$s \\
syndrome processing time & $t_{\rm sp}$ & 5 $\mu$s \\
{logical gate rate} & $1/(2w \tau_g + 2 t_{\rm m} + t_{\rm sp})$ & 8 kHz \\
2-p-bit gate error & $\gamma_2 \simeq \epsilon_s + P_{\bar{n}}$ & $10^{-4}$ \\
1-p-bit gate error & $\gamma_1$  & $10^{-3}$ \\
measurement error & $\gamma_m$   & $10^{-3}$ \\
memory error & $\epsilon = 1/Q$  & $10^{-6}$ \\
recovery crash probability & $p \simeq$ & $10^{-10}$ \\
n. logical gates & $\simeq 1/(bp) \simeq$ & $10^9$
\end{tabular}}
\end{table}

\vspace*{4pt}   
\begin{table}[hb]
\tcaption{Physical parameters of the computer. The list is in the
order of the argument presented in the text. The bold parameters
are the main ones used in table 1. The example values are given
for the Cd$^+$ ion. Parameters for Ca$^+$ at the same gate
rate give smaller laser power and a more convenient wavelength, but
a more difficult hyperfine repumping.}
\centerline{\footnotesize\smalllineskip
\begin{tabular}{lll}
Optical \\
linewidth & $\Gamma$ & $2\pi \times 44$ MHz  \\
collection efficiency & $\varepsilon$ & 0.02 \\
mean counts per ion & $\bar{c}$ & 10 \\
~P(0 or 1 count) & $(1+\bar{c})e^{-\bar{c}}$ & 0.0005 \\
~{\bf measurement time} & $t_{\rm m} = 4 \bar{c}/\varepsilon \Gamma$ & 7.2 $\mu$s \\
wavelength & $\lambda$ & 214 nm \\
~saturation intensity & $I_0 = 4 \pi^2 \Gamma \hbar c / 3 \lambda^3$ & 11700 Wm$^{-2}$ \\
mass number & $A = m/{\rm u}$ & 111 \\
~recoil frequency & $R = h/2 A {\rm u} \lambda^2$ & 39 kHz \\
fine structure & $\omega_F$ & $2\pi \times 74$ THz \\
~scattered photons at local minimum & $P_0 \simeq 2 \sqrt{2} \pi {\Gamma}/{\omega_F}$ & $5.3\times 10^{-6}$ \\
{\bf infidelity from photon scattering} & $\epsilon_s$  & $4\times 10^{-5}$ \\
{\bf phase-gate time} & $\tau_p$ & 0.5 $\mu$s \\
~stretch mode frequency & $\nu_{\rm str} \simeq 4/\tau_p$ & 8 MHz \\
~c.o.m. mode frequency & $\nu_{\rm com}=\nu_{\rm str}/\sqrt{3}$ & 4.6 MHz \\
~stretch-mode Lamb Dicke param. & $\eta = \sqrt{R/\nu_{\rm str}}$ & 0.07 \\
~carrier Raman Rabi frequency & $\Omega_{\rm R} = \pi / \eta \tau_p$ & $2 \pi \times 14$ MHz \\
~laser intensity for $P_0$ & $I_{P0} = 6 \omega_F(3\sqrt{2}-4) \Omega_R I_0 / \Gamma^2$ & 9.2 mW$/\mu$m$^2$\\
~intensity per laser beam & $I = I_{P0} P_0/(\eta \epsilon_s - \eta^2 P_0)$ & 18 mW$/\mu$m$^2$ \\
beam diameter & $2r$ & 4 $\mu$m \\
~power per beam & $\pi r^2 I$ & 220 mW \\
~total laser power & $ 2 N_P \pi r^2 I$ & 440 W \\
\hline
Trapping: axial (d.c.) \\
nearest distance to electrode & $\rho$ & 10 $\mu$m \\
d.c. electric field at electrode & $E_{\rm max}$ & 200 V$/\mu$m \\
octopole geometric factor & $\mu_8$ & 0.02 \\
~c.o.m. frequency at split & $\nu_{\rm spl} = \frac{840}{2 \pi \sqrt{A}}\left(\frac{\mu_8 E_{\rm max}}{\rho^3}\right)^{3/10}$ & 15 MHz \\
\hline
Trapping: radial (r.f.) \\
Mathieu $q$-parameter & $q_r$ & 0.3 \\
r.f. quadrupole geometric factor & $\mu_4$ & 0.15 \\
r.f. electric field amplitude & $E_{\rm rf}$ & 100 V$/\mu$m \\
~r.m.s. voltage & $V_{\rm rms} =  \mu_4 \rho E_{\rm rf}  /  \sqrt{2}$ & 106 V \\
~radial secular freq. & $\nu_r = \frac{1}{2\pi}\left( \frac{q_r e}{2 A u} \frac{\mu_4 E_{\rm rf}}{\rho}\right)^{1/2}$ & 70 MHz \\
~r.f. frequency & $\Omega/2 \pi = 2 \sqrt{2} \nu_r / q_r $ &  660 MHz \\
\hline
Electrical architecture \\
~total area & $50 N \rho^2$ & 0.3 cm$^2$ \\
~n. d.c. electrodes & $30 N_P + 20 N$ & 160,000 \\
~electrode density &       & 490,000 cm$^{-2}$ \\
~capacitance per p-bit & $C=20 \rho \epsilon_0$ & 0.002 pF \\
loss tangent & $\tan \delta$ &  0.0005 \\
~total r.f. power dissipated & $N V_{\rm rms}^2 \Omega C \tan \delta$ & 24 mW \\
\hline
Thermal\\
mean vibration number & $\bar{n}$ & 0.5 \\
~{\bf Lamb Dicke gate error} & $P_{\bar{n}} = 0.3 \pi^2 \eta^4 \bar{n}(\bar{n}+1)$ & $5 \times 10^{-5}$ \\
~cooling time & $\tau_{\rm cool} = 1/\bar{n} \nu_{\rm com}$ & 0.4 $\mu$s \\
heating rate in gate zone& $d\bar{n}/{dt}$ & 1 ms$^{-1}$ \\
~heating during phase gate & $\Delta \bar{n} = \tau_p d \bar{n}/{dt}$ & $\sim 0.0005$ \\
\end{tabular}}
\end{table}

The computer is made from two main elements: an array of linear
r.f. Paul traps on an `ion chip' (IC) held in high vacuum, with
control circuitry for the trap electrodes integrated on the chip,
and an optical system outside the vacuum chamber. The concept is
broadly as proposed by Kielpinski et al. \cite{02Kielpinski}.
Qubits are stored in the hyperfine structure of ions; these are
held in an array of ion traps on a chip in high vacuum, moved
around by control voltages on the multiple electrodes, manipulated
by coherently driven Raman transitions, and measured by
laser-stimulated fluorescence. The IC contains $160,000$
electrodes of dimensions of order 1 $\mu$m, individually
controllable at switching rates of order 50 MHz and voltages of
order 100 V by the classical control circuitry integrated on the
IC.

The optical system must deliver 990 laser beam pairs to the IC,
aligned to $\sim 1 \,\mu$m with a power of 10 to 200 mW per beam
(depending on the ion species and pointing stability). Each
pair should have a precise
frequency difference of order 10 GHz, or else the beams may each
be frequency-modulated, depending on details of how the gates are
implemented. This r.f. oscillation must be controlled by
phase-locked high-quality circuitry similar to that used in
nuclear magnetic resonance spectrometers, delivering pulses of
duration 0.1 to 1 $\mu$s. The laser beams should be switched with
very high extinction ratio on the same timescale.

Major savings in the construction scale and the power requirements
would be made if the optical system could be integrated onto a
second chip. This is an important aim for future technology, but
even if it could not be done then one could still realise the
optical system, by duplicating current techniques such as
solid-state lasers and acousto-optic modulators a thousand-fold.
This is discussed further below.

For each QEC of the computer, around 1500 measurements are
required at once, each involving 2 ions. Optical detection could
be integrated onto either the IC or an optical chip, or else a
separate detection chip is used, having at least 1500 pixels which
can be read out individually in a time scale of order 1 $\mu$s.

The large-scale parameter values, such as the total number of ions
and laser beam pairs, are dictated by the requirement to perform
QEC repetitively, and the decision of how large a machine to
attempt. For the sake of having a concrete example that would
comfortably out-perform any classical computer on suitably chosen
algorithms, I choose the benchmark of $10^9$ logical operations on
300 logical qubits. That is, the computer is designed to handle an
algorithm of this size with a substantial probability of
success\footnote{Algorithms, both classical and quantum, typically
require many more gates than bits; these values would be
appropriate if the number of gates scales as $40 n^3$, for
example.}. A single 300-qubit computer could not factorize numbers
of interest to cryptography, but quantum algorithms to study
eigenvalue problems, spin systems, and quantum chemistry are
predicted to out-perform classical computers once 50 to 100 qubits
are available\cite{99Abrams,04Porras,05Aspuru}.

Various types of QEC encoding are possible; in particular there is
a trade-off between the noise level that can be tolerated and the
number of physical qubits required \cite{03Steane,04Knill2}. I choose to
design for gate noise at the level $10^{-4}$, and adopt the most
efficient encoding that can achieve the required noise-tolerance.
A more full statement of the error model is given in section 2.
The result of these choices is a large block-code encoding using
the $[[n,k,d]]=[[127, 29, 15]]$ CSS code based on a classical BCH
code \cite{96Calderbank,96Steane2,03Steane}. This performs very
well under random errors. In order to enhance the robustness to an
important type of systematic error, namely correlated phase error,
such as is caused by joint energy level shifts of adjacent ions,
or drift of a local oscillator phase, a low-level encoding using a
two-qubit `decoherence free subspace' is concatenated with the
large code \cite{BkNielsen,04Roos,01Kielpinski}. After the QEC is
taken into account, I find a logical gate rate of approximately
10~kHz, limited by the physical gate rate, so that it
would take about 30 hours to complete the $O(10^9)$ logic
operations which are in principle available. This might appear
`slow' (though of course it still out-performs any classical
computer on suitably chosen algorithms), but the logical gate rate
of a QC will necessarily be slow compared to that of a good
classical processor, first because the QC relies on very precise
classical control circuitry, and secondly because a non-negligible
amount of classical processing must take place in each logical
step, in order to interpret the 78-bit syndromes of the BCH code.
I have allowed 5 $\mu$s for this classical processing. The reason
for this choice is that it is long enough to appear reasonable,
and short enough to have almost negligible impact on the logical
gate rate for the computer under discussion. The main limits on
speed come from the need to avoid photon scattering when the
quantum gates are driven by optical pulses, and the time taken to
move ions around the channels in the chip; I discuss faster
operation below.

\subsection{Feasibility}
\noindent
The IC has vacuum channels of width $20\,\mu$m,
electrodes of width around $1\,\mu$m, voltages in the region 10 to
100 V, and an r.f. frequency for the Paul traps of 660 MHz. All
these values are close to those already implemented in single ion
traps or ion trap arrays \cite{95Jefferts,04Blain,04Barrett,04Madsen,04Home}.
The major new ingredient is the large number of electrodes ($\sim
160,000$) and parallel operations ($\sim 1000$ are required to
allow fault-tolerant QEC at the assumed noise levels). This
implies that a large amount of control circuitry must be
integrated onto the IC because there is not room for sufficient
wires to circuitry located elsewhere. Although this is a
considerable design problem, the size and speed (switching times
of order 10 ns) of these circuits are within the capability of
current fabrication techniques. The substrate could for example be
silicon.

The measurement and quantum gate techniques are precisely those
already demonstrated in studies such as
\cite{03Leibfried,04Barrett,04Riebe,05Brickman,05Haffner,05Haljan,05Leibfried,04Chiaverini};
I have merely picked some reasonable numbers for detection
efficiency, and assumed sufficient laser power to enable the gate
to be operated by a Raman process sufficiently far detuned that
Raman scattering is below $10^{-4}$ per gate.
I also assumed technical problems such as laser intensity noise
and mechanical jitter on optical paths can be reduced sufficiently
so that their contribution to errors per gate is below $10^{-4}$.
This would require the path difference of each Raman beam pair to
be stable to $\sim$ a few nm during $\tau_p$ (defined in table 2),
but it can vary between gates by much larger amounts. Electrical
noise, which may include cross-talk in the IC electronics, must be
small enough to allow the ion locations to be similarly stable
during the phase-gate time.

Placing the optical system outside the vacuum chamber would reduce
several aspects of the fabrication problem. The high-quality
imaging optics and interferometrically-stable mounting which are
required to focus the laser beams onto the IC are not especially
demanding compared to high-end optical instrumentation currently
available. If each laser beam diameter at the IC is $4\, \mu$m in
order to make the required numerical aperture and alignment
precision reasonably straightforward, then the total laser power
required for $\sim 1000$ parallel operations is
from some tens to several hundred Watts, depending on speed and,
to a lesser extent, on ion species. It is
important therefore to prevent these laser beams impinging on the
IC. This is possible either with a planar design and laser beams
propagating parallel to the IC surface, or by fabricating the IC
with holes in it to allow the beams to pass through. These holes
would only use up a small fraction of the chip area available for
control circuitry. A smaller beam diameter might be preferred in
order to reduce the power requirement. If instead the optical chip
were inside the vacuum chamber, then the focussing elements would
be integrated onto it and it would be close to the IC (e.g. a few
mm away). This construction would exclude air pressure
fluctuations and would allow higher rigidity, hence allowing
smaller laser beam diameter and power.

Scattering during readout must also be avoided. The laser beams
used for readout have intensity of order the saturation intensity
$I_0$, i.e. some $10^6$ times smaller than the Raman beam pairs. A
fluorescing ion emits a total power of order $\Gamma h c/2\lambda
\simeq 10^{-10} W$, while each readout laser beam power is of
order $\pi r^2 I_0 \simeq 1\;\mu$W. To make the scattered
background considerably smaller than the fluorescence rate, we
require less than $10^{-5}$ of the readout beam power to be
scattered.

The total laser power required for the Raman beams is also
feasible, though since laser sources are inefficient it would
require careful power handling to extract heat. The major problem
is to control 1000 laser beam pairs in high precision pulses of
duration of order $1\,\mu$s. To produce the required pulses
current experiments use acousto-optic and electro-optic
modulators. These are of centimetre dimensions and dissipate 1
Watt of r.f. drive power each. An array of thousands of such
devices maintaining the required stability would be a formidable,
though perhaps possible, engineering challenge. However, steering
mirror arrays fabricated by MEMS can already achieve switching
times of order $10\,\mu$s, and sidebands can be impressed onto
semiconductor diode laser output by low-power r.f. modulation of
the drive current. It is reasonable to expect that these or
related techniques will make the optical system feasible.

The physical gate rate is found to be of order 1 MHz, and this
places the main limitation on the logical gate rate. However, given
that the quantum gates rely on precise r.f. control pulses, it is important
that we do not assume too fast an operation of the classical control
circuitry, otherwise that would itself become the major constraint on
feasibility.

\section{Methods}

The QC is designed to freely manipulate 290 logical qubits, and to do this it
uses a further 58 logical qubits in two blocks as workspace in
fault-tolerant logical operation circuits
\cite{96Shor,BkNielsen,98Gottesman1,99Gottesman2,05Steane}. There
are two layers of encoding: first each pair of physical qubits
stores a single logical bit by the `decoherence free' encoding
$\ket{0} \rightarrow \ket{01}$, $\ket{1} \rightarrow \ket{10}$,
which protects it from joint phase
noise\cite{BkNielsen,04Roos2,01Kielpinski,02Kielpinski}. For
brevity I refer to these pair-encoded logical bits as `p-bits'.
The p-bits are then encoded in a large block code, the
$[[n,k,d]]=[[127, 29, 15]]$ CSS code based on a classical BCH code
\cite{96Calderbank,96Steane2,03Steane}. Each p-bit, or else a
substantial fraction of the p-bits, is accompanied by a further
ion of another species in order to allow sympathetic cooling.

Let the quantum memory quality factor $Q$ be defined as the ratio
of the decoherence time of a p-bit to the time required to
complete a controlled operation on a pair of p-bits which were
initially separated by a distance of order 10 p-bits. Let
$\gamma_2$ be the imprecision (error probability) of a controlled
phase gate between two p-bits which were initially separated by of
order 10 p-bits. $\gamma_1, \gamma_m$ are the error probability of
a Hadamard (or equivalent) rotation of a single p-bit, and of a
measurement of a p-bit, respectively. The reason to specify a
distance of 10 p-bits comes from a study of typical distances in
QEC recovery circuits \cite{02Steane2}, I will return to this
point below.

The estimate in table 1 of the `crash' probability $p$ per
recovery of one block is calculated using the formulae in
\cite{03Steane}. The recovery networks have many more 2-bit gates
(these can be controlled-not or controlled-phase) than 1-bit gates
or measurements, therefore the errors $\gamma_1,\, \gamma_m$ can
be substantially larger than $\gamma_2$ without much effect on
$p$.

To establish the speed of the computer, first consider measurement
of physical qubits. A single recovery involves 2
measurements (1 to verify the ancilla, 1 to acquire the syndrome).
Therefore twice the measurement time is a lower bound on the logical gate time,
unless the computer size is increased in order to prepare further
ancillas in the background so that they are ready when needed.

The hyperfine state of each ion is inferred from laser-stimulated
fluorescence. I assume a detector counts photons, and I set a
threshold of 2 counts, so that the measurement error is estimated
in table 2 as the probability that a fluorescing ion only gives 0
or 1 counts. For this probability to fall below $10^{-3}$ we
require the mean count to be 10 or more. A reasonable value for
combined collection and detection efficiency is $0.02$, and
therefore\footnote{The excited fraction for a saturated ion is
between $1/4$ and $1/2$, depending on the gross structure.} the
measurement time is $t_{\rm m} \simeq 2000/\Gamma \simeq 10\;\mu$s.

Another basic time-scale is set by laser power constraints and
the need to avoid photon
scattering during physical gate operations. The 2-p-bit controlled
gate can be achieved either by a controlled-phase gate between one
ion from each pair, or by a joint gate on all four ions similar to
that described in \cite{02Kielpinski,99Sorensen}. The physics of
the spin-flip and phase-flip gates described in
\cite{99Sorensen,03Leibfried} respectively is similar. The former
may be preferred if the qubit transition is first-order
insensitive to magnetic field changes \cite{03Wineland,05Haljan};
the latter may be preferred otherwise. The physical mechanism of
these and other gates for trapped ions have in common the use of
laser-excitation of the motion, and the fact that approximately
$P_{0}/\eta$ photons are scattered per ion during the gate, where $P_0$ is
the number which would be scattered during a
$\pi$-pulse on the carrier (i.e. non motion-changing) excitation,
and $\eta$ is the Lamb-Dicke parameter. Decoherence of the hyperfine state
is caused by Raman (not Rayleigh) scattering \cite{05Ozeri}.
The total (Raman and Rayleigh) scattering also heats and decoheres the
motion, which is important during a 2-qubit gate when the qubits and motion
are entangled. To estimate the latter, note that one photon recoil
$\hbar k$ at time $t$ causes a change in coherent state
parameter $\Delta \alpha=i \eta \exp(i 2 \pi \nu_{\rm str} t)$. This
changes the orbit area and hence accumulated phase by an
amount of order $\eta$. After $N_s$ scattering events at random
times the resulting infidelity is of order ($N_s \eta^2/2$).
The total infidelity due to scattering, $\epsilon_s$, is the sum of
this and the half the number of Raman-scattering events during
a gate\footnote{This assumes that population `leaking' away
by Raman scattering to hyperfine Zeeman sublevels outside
the computational basis can be reclaimed
at little cost, for example by optical pumping. This is non-trivial
however and may favour ions with small $F$ and large hyperfine splitting.}.
To obtain the latter, note that $\epsilon_s < 10^{-4}$
requires a detuning $\Delta$ from dipole-allowed transitions
of the order of or larger than the fine structure. We first
consider the local minimum
in $P_0(\Delta)$ when the detuning is in between the fine
structure components,
near $\Delta = (\sqrt{2}-1)\omega_F$, see \cite{03Wineland}.
At this detuning, the total scattering is dominated by
Raman scattering, and for a `clock' ($M=0$--$0$) transition,
assuming linear polarizations in order to reduce differential
Stark shifts, the local minimum value
is $P_0 \simeq 2 \sqrt{2} \pi \Gamma/\omega_F$ per ion. This suggests
that large fine structure splitting $\omega_F$, and therefore
a heavier ion, is advantageous. However, for $|\Delta| \gg \omega_F$
all the usual ions (groups IIA, IIB and Yb) can achieve
low enough $\epsilon_s$, so other considerations,
such as required laser wavelength and power, and hyperfine repumping
are more important.
Let $I_{P0}$ be the laser intensity which would be required to obtain the
desired gate rate when $\Delta$ is at the local minimum. Then at the actual
operating point $|\Delta| > \omega_F$, the Raman scattering per
gate is smaller than $2 P_0/\eta$ by the ratio $I_{P0}/I$
where $I$ is the laser intensity\footnote{We ignore other
gross structure levels, which is valid as long as $\epsilon_s$
is not too small.}, while the total scattering per gate $N_s \simeq 2 P_0/\eta$.
Therefore the total infidelity due to photon
scattering from both ions is $\epsilon_s \simeq (I_{P0}/I + \eta^2)P_0/\eta$.
In table 2 we specify the required $\epsilon_s$ and invert this into
a formula for $I$.

By using well-chosen pulse design, the gate rate can
approach and even exceed the vibrational frequency
\cite{03GarciaRipoll}, but this exacerbates photon scattering.
In order not to assume too great a
reliance on advanced techniques, I take the inverse gate time
$1/\tau_p$ to be of the order of the motional frequencies (centre
of mass and stretch modes) of one pair of ions, $\tau_p = 4/\nu_{\rm str}$.
Usually $\eta P_0 \ll \epsilon_s$ so the $\eta^2$ (heating) term in $\epsilon_s$
can be dropped, and then
\beq
\tau_p \propto (I\lambda \epsilon_s/m)^{-1/2}.
\eeq
This formula summarises the way atomic parameters and
available laser intensity constrain
the gate time for a pair of neighbouring ions.
To make this
time contribute similarly to the measurement time for each logical
gate, one would choose (see table 1) $\tau_p \simeq t_{\rm m} / w
\simeq 0.2\;\mu$s. However, the laser power requirements are
severe, and therefore I adopt here $\tau_p \simeq 0.5\;\mu$s in order
to save power without much cost in overall speed
(this gives $\eta=0.07$ for Cd, $\eta=0.06$ for Ca). With the
further contribution of the time for moving ions around,
this implies that the logical gate rate of this QC is limited by the
physical gate rate, not the measurement time.

A complete gate operation between p-bits initially at different
locations involves 5 steps: split, move, recombine, cool, operate.
The `split' is the process of pulling apart two ions which were
close together (in order to separate one p-bit from another
p-bit), `recombine' is the reverse process. I allow a time of
$2/\nu_{\rm spl}$ for the split and recombine stages, where
$\nu_{\rm spl}$ is the centre of mass frequency of a pair of ions
at the moment during the split when the potential barrier between
them just appears, so that they are in a quartic rather than
quadratic potential. This frequency can be made large by using
large voltages on the electrodes, but there is a limit set by
electrical breakdown. This is discussed in \cite{04Home} and leads
to the formula given in table 2 in terms of a geometric factor
$\mu_8$ and the maximum allowed electric field at the electrode
surfaces $E_{\rm max}$. Micro-fabricated ion trap structures of
similar dimensions to those considered here have been found to
operate at electric fields up to $10^{9}$ V/m \cite{04Cruz}, which
is the region where field emission can occur. To be cautious I
take the maximum allowed d.c. electric field to be $2 \times 10^8$
V/m. The split time scales almost linearly with the distance scale
of the electrodes, parametrised by the nearest distance $\rho$
from a trapped ion to an electrode surface. The value of $\rho$ is
limited by fabrication ease and by heating problems.

I model heating of trapped ions as due to electric field noise at
the ions, in which case the heating rate is given by $d\bar{n}/dt
= e^2 S(\nu)/4 m h \nu$ where $\bar{n}$ is the mean vibrational
quantum number and $S$ is the spectral density of electric field
fluctuations (dimensions (V/m)$^2$Hz$^{-1}$). The results of a
number of experimental studies of heating rates in small ion
traps, summarised in \cite{04Deslauriers}, can be fitted by the
empirical formula $S \simeq a / \rho^4$ where $a = 10^{-26 \pm 1}$
(Vm)$^2$Hz$^{-1}$\footnote{A model to explain the noise level and
the $\rho^4$ dependance is put forward in \cite{00Turchette}, the
large uncertainty in $a$ is due to the fact that it may depend on
surface quality and ion species as well as other factors.}. If we
require that during the split time this heating results in $\Delta
\bar{n} \le 1$, then we obtain a lower bound $\rho \ge 5\;\mu$m
when $\mu_8 E_{\rm max} = 4 \times 10^6$ V/m. This is for
electrodes at room temperature; the heating could be reduced by
cooling the electrodes.

A lower bound $\rho \gg \lambda$ applies to the gate zone
(i.e. where quantum logic-gate laser pulses are applied),
coming from the need to allow laser
beams to be kept clear of the electrode structures. This also results
in $\rho$ of order microns.

Next, consider moving the ions once they have been split. Each
p-bit pair is moved as a single entity. The required networks of
operations are chiefly those which create and verify ancillas. It
was shown in \cite{02Steane2} that if the qubits are laid out
along a line, then the mean separation of bits which must have a
gate between them, averaged over all the gates in the network, is
$22$ bits. This gives the order of magnitude of the distance over
which ions will have to be moved. It is advantageous to arrange
the ions of each ancilla in a rectangular array rather than a
line, which will reduce the average communication distance. I take
the total movement time to be $10/\nu_r$ where $\nu_r$ is the
radial confinement frequency calculated for a Paul trap having
distance scale $\rho$, Mathieu $q$ parameter $0.3$ and maximum
r.f. electric field $10^8$ V/m, see table 2.

I adopt $\rho = 10\;\mu$m for splitting and transport regions of
the chip in order to obtain a high speed while keeping the heating
rate acceptably small, so that once ions arrive at a gate region
they can be laser-cooled rapidly. To reduce heating of ions during
the controlled logic-gate laser pulse where it would result in infidelity,
the gate should be implemented in a zone slightly displaced from
the splitting/recombining zone, with electrode surfaces further
away by a factor 2 to 10.

The laser cooling requirement is set by the impact of non-zero
temperature on gate fidelity. This was studied in
\cite{00Sorensen1} for a gate in the general class of those which
are insensitive to thermal effects at first order in the
Lamb-Dicke parameter, but sensitive to second order. The results
give a reasonable guide for present purposes, where I assume that
a gate in this class is to be implemented. It is found that to
obtain an infidelity from thermal effects of $5 \times 10^{-5}$,
it is sufficient to cool to a mean vibrational quantum number
$\bar{n} = 0.5$ when $\eta = 0.07$. I assume that the motional
heating which took place during the split and move only resulted
in a few phonons' worth of heating. This and the fact that the
required $\bar{n}$ is not small compared to 1 allows the cooling
time to be small, close to the inverse of the motional frequency.

I find that each of the 5 stages of a physical gate takes a
similar time. The total time required is given in table 1, it is
$1.2\,\mu$s for Cd, $1.08\,\mu$s for Ca.

This completes the discussion of the parameters giving the speed
and noise tolerance of the computer. It remains to comment on the
IC and optical sources.

The IC has of order 160,000 electrodes, allowing for 30 electrodes
for each zone where gate operations take place, and a further 20
per p-bit for all the bits. Electrode voltage ramp times are of
order 10 ns, with 990 separate zones of the IC to be activated in
parallel. To allow on-chip logic and digital-to-analogue
converters it is likely that silicon is the best substrate, though
exploration of materials questions has only just begun. I assume
that metal electrodes on a $\sim 10\;\mu$m thick layer of SiO$_2$
can be modelled roughly as capacitors with an r.f. loss tangent of
order $0.0005$. The distance scale $\rho$ and the values given in
table 2 for the radial confinement then imply the r.f. power
dissipation is small (assuming ohmic heating in the conductors
is negligible). 

The laser beams used for gate operations are detuned by around
the fine structure splitting, and this means that in order to obtain
the desired gate speed they must be intense, with $I \simeq
18\;{\rm mW}/\mu{\rm m}^2$ for Cd$^+$. To ease alignment the laser
beam waists should not be too small. I allow a beam diameter $4
\;\mu$m. For Cd$^+$ this is $19 \lambda$, for Ca$^+$ it is $10
\lambda$, and therefore it is readily achievable in either case with
high-quality optics. To reduce alignment errors and to prevent stray
light from hitting electrodes it may be useful to use larger beams
and pass them all through a reflective mask with $4\,\mu$m holes in
it, which is then imaged onto the computer chip. In any case a
uniformly illuminated $4\;\mu$m spot represents 220 mW  per laser
beam at the required intensity. This suggests the total laser power
is $2 \times 0.22 \times 990 \simeq 440$ W for
Cd$^+$ (100 W for Ca$^+$). In order to reduce the total
instantaneous power one could separate in time the
pulses on different ion pairs; this does not slow the computer
when $\tau_p < \tau_g$.

\section{Faster, larger, noisier}
\noindent
I conclude with a brief discussion of improvements and other
approaches, towards faster operation and larger algorithm
capability.

The logical gate rate could be increased without any change in the
hardware apart from scaling up, by devoting more resources to
preparing ancillas. There is up to a factor $\simeq 2 w \simeq 100$
gain in speed available this way (i.e. until the recovery time
is similar to the gate time), at a cost of the same factor
increase in the number of ions, traps and laser beams.

Scaling up the number of physical bits, without a change in the
number of logical bits, could instead be used to allow a higher
noise threshold, by using a different QEC encoding. The threshold
increases slowly with the computer size however; a very rough rule
of thumb which summarises results in \cite{03Steane} is $N \propto
\gamma^{2.5}$ for $\gamma < 0.003$ where $\gamma$ is the physical
gate error; higher noise is also possible with a similar cost
\cite{04Reichardt,04Knill2}.

To reduce the measurement time some further ions could reside in
each measurement zone on the chip, and when a given p-bit is to be
measured, it is first coupled by controlled-not to these further
ions and then all the ions in the set are simultaneously measured
and a classical majority vote is taken to determine the result
\cite{98Stevens}. A speed-up is obtained as long as the
controlled-not operations are fast compared to the measurement
time.

To reduce the phase-gate and cooling time one could adopt fast
pulse techniques as for example in \cite{03GarciaRipoll}, and
increase the trap frequencies. This would require larger
laser detuning and power to combat photon scattering.
Once the measurement and phase gate time is reduced, it becomes
worthwhile to use smaller electrode distance $\rho$ in order to
reduce the split and move time.

These improvements could gain several orders of magnitude in
speed, or they could be used to make the machine more robust and
therefore relax the engineering requirements. To further increase
the gate rate it would ultimately be necessary to abandon optical
methods to control and measure the ions, and adopt electronic
methods instead. For example, one might transfer quantum
information directly between an ion and an electron in a closely
situated electrode \cite{04Tian}. However, one would then lose
some very attractive features of optical control, and the
electrons would be subject to a much larger and more complex set
of processes which lead to decoherence than the quasi-free ions.
The advantages of optical methods are chiefly that photons do not
directly influence one another, they are insensitive to
electromagnetic field noise, and their excitation energies are far
above the thermal energy at room temperature, so they can be
switched off completely.

\nonumsection{Acknowledgements}
\noindent
This work was supported by the European Union through the 
IST/FET/QIPC project ``QGATES" and by the National Security Agency
(NSA) and Advanced Research and Development Activity (ARDA)
(P-43513-PH-QCO-02107-1).

\bibliographystyle{unsrt}
\bibliography{myrefs}

\end{document}